\documentstyle[a4,spie]{article}

\begin{document}

\title{Optimal conditions for light pulses coherence transformation
in thin nonlinear media}

\author{M. Karelin, A. Lazaruk 
\thanks{Institute of Physics, National Academy of Sciences
F. Skaryna Ave. 70, Minsk, 220072, Belarus. 
E-mail: karelin@ifanbel.bas-net.by, lazaruk@ifanbel.bas-net.by}}
\date{}

\maketitle

\section*{ABSTRACT}
Via solution of appropriate variational problem it is shown that 
light beams with Gaussian spatial profile and sufficiently short 
duration provide maximal destruction of global coherence under 
nonlinear self-modulation.

{\bf Keywords:} nonlinear phase self-modulation, controllable 
coherence degradation.

\section{INTRODUCTION}

Earlier it was demonstrated \cite{KarelinLazaruk}, 
that self-modulation of light fields 
in a nonlinear medium can serve as a convenient 
tool for controllable coherence manipulation. Process of local 
interaction of coherent incident field 
$E_{in}({\bf r},t)={\cal E}({\bf r})e(t)$ with ``optically 
thin'' nonlinear layer, being described by Raman-Nath approximation, 
causes different nonstationary phase shifts
$$
E_{out}({\bf r},t)=E_{in}({\bf r},t)
\exp \left\{
i\Phi(|E_{in}({\bf r},t)|^2)
\right\}, 
\eqno(1)
$$
where the phase is determined by media parameters and intensity of 
input light in every particular point ${\bf r}$ of a beam 
cross-section. The resulting degradation of spatial coherence can be 
treated as a decay of initially single-mode (but 
nonuniform) radiation into a number of mutually incoherent, 
orthogonal modes. Such a process imitates the action of 
moving phase diffuser and can in principle be used for speckle-noise 
reduction in experiments with short light pulses.

The main characteristic of output field in discussed process is 
an overall degree of coherence
$$
\mu={1 \over U^2}\int dt_1 \int dt_2 \big|K(t_1,t_2) \big|^2,
\eqno(2)
$$
where
$$
K(t_1,t_2) = \int d^2r\, E_{out}({\bf r},t_1) E^*_{out}({\bf r},t_2)
\eqno(3)
$$
is a spatially averaged temporal correlation function, and
$$
U=\int d^2r \int dt |E_{in}({\bf r},t)|^2 = \int dt\, K(t,t)
$$
--- pulse energy (nonlinear medium is supposed to be absorbtionless). 
The value (2) determines the contrast of all 
interference phenomena (including speckles) and it is closely 
connected with coherent-mode structure of optical 
fields. Modal treatment is an analogue of Karhunen-Lo\'eve 
transformation, and the main parameter of such approach 
--- effective number of modes --- is $N_{eff}=1/\mu$ 
(see Ref. 2, 3 for further details).

\section{OPTIMISATION OVER SPATIAL BEAM DISTRIBUTION}

The consideration in paper \cite{KarelinLazaruk} was carried out 
for the model of 
cubic nonlinearity with exponential relaxation and 
speckled input field. Maximal coherence destruction there is 
achieved in the ultimate case of inertial interaction with 
infinite memory
$$
\Phi(|E_{in}({\bf r},t)|^2)=
\eta \int^t_{-\infty} dt |E_{in}({\bf r},t)|^2 .
\eqno(4)
$$
For this limit the value (2) does not depend on temporal shape of 
input pulse, and every initial field is equivalent to 
rectangular pulse of duration $T$:
$$
E_{in}({\bf r},t)=\left\{ 
	{{\cal E}(r)/\sqrt{T} , \; 0<t\le T \atop
              0, \; {\it elsewhere}}
\right.
$$

The interaction with infinite memory provides maximal total phase 
shift, so one can expect, that the case (4) will 
result in maximal coherence destruction for any input beam. Hence 
the main aim of the present analysis is to optimise 
the transformation (1), (4) over possible spatial distributions 
of initial field ${\cal E}({\bf r})$, what can be done via solving 
variational problem on minimum of double integral 
$$
\mu U^2=2\int d^2 r_1 \int d^2 r_2 \, I({\bf r}_1) I({\bf r}_2)
\frac{1-\cos(\eta T[I({\bf r}_1)-I({\bf r}_2)])}
     {\eta^2 T^2 [I({\bf r}_1)-I({\bf r}_2)]^2}
\eqno(5)
$$
under additional constrain of constant energy
$$
\int d^2 r I({\bf r})=U,
\eqno(6)
$$
where $I({\bf r})=|{\cal E}({\bf r})|^2$ --- input intensity profile.

So far the above two functionals do depend on incident intensity 
only, the task can be simplified by transition to 
integration over beam intensity with proper introduction of 
quasi-distribution function $P(I)$: 
$$
K(t_1,t_2)={1 \over T}\int^{I_0}_0 dI \, P(I) I \exp \{i \eta I(t_1 - t_2)\},
\qquad
(0 \le t_1, t_2 \le T),
\eqno(3a)
$$
$$
\mu U^2=2\int^{I_0}_0 dI_1 \int^{I_0}_0 dI_2 \, P(I_1) P(I_2) I_1 I_2
\frac{1-\cos(\eta T[I_1-I_2])}{\eta^2 T^2 (I_1 - I_2)^2},
\eqno(5a)
$$
$$
\int^{I_0}_0 dI \, P(I) \, I = U.
\eqno(6a)
$$
Function $P(I)$ has a meaning of measure of contribution into these integrals 
due to field points of given intensity. In 
particular important case of axially symmetric and monotonous 
dependence $I(r)$ it is easy to see that 
\hbox{$P(I)=2\pi r(I) |\partial r/\partial I|$}. Here additional 
condition of finite maximal intensity $I_0$ is implied. It is reasonable 
for any real field distribution (but it does not the 
case for speckle model).

At sufficiently high values of $\eta$ one can do further simplification 
on the base of equality 
$$
(1-\cos \eta x)/\eta^2x^2 \approx \delta(x) \pi/\eta,
$$
and then (5a) takes the form
$$
\mu U^2 = \frac{2\pi}{\eta T}\int^{I_0}_0 dI\, P^2(I)\,I^2.
\eqno(5b)
$$
Then it is readily seen that maximal coherence destruction is achieved when
$$
P_{MAX}(I)={U \over I_0}{1 \over I},
$$	
what in coordinate representation takes the form of Gaussian beam
$$
I_{MAX}(r)=I_0 \exp(-\pi r^2 I_0/U),
\eqno(7)
$$
and appropriate overall degree of coherence is 
$$
\mu = \frac{2\pi}{\eta T I_0}.
\eqno(8)
$$

\section{DISCUSSION}

The derived profile of input beam has some unique features, that makes 
it especially attractive for nonlinear 
speckle-noise reduction. First, such fields are easy to generate, 
to control and to operate. 

Second, in bulk a layer of nonlinear media
\cite{KarelinLazaruk} diffractional mixing 
diminish the resulting decoherence of a field. 
Gaussian beam has absolutely minimal diffractional divergence and 
consequently it allows to use thicker layers, 
increasing the efficiency of coherence transformation.

At last, it should be noted that according to paper
\cite{LazarukKarelin} zones of 
identical intensity (for (7) these are concentric 
circles) belong to one mode or, in other words, produce coherent 
radiation. In order to generate output light with 
desired structure of spatial coherence function one just need 
to mix the field after nonlinear media on suitable 
stationary diffuser.

\section*{ACKNOWLEDGEMENTS}

This work was supported by the Belarusian Foundation for Basic 
Research under grant No. F97-253.

\end{document}